# Data Acquisition and Control System for Broad-band Microwave Reflectometry on EAST

Fei Wen, Tao Zhang, Shoubiao Zhang, Defeng Kong, Yuming Wang, Xiang Han, Hao Qu, and Xiang Gao

*Abstract*–Microwave reflectometry is a non-intrusive plasma diagnostic tool which is widely applied in many fusion devices. In 2014, the microwave reflectometry on Experimental Advanced Superconducting Tokamak (EAST) had been upgraded to measure plasma density profile and fluctuation, which covered the frequency range of Q-band (32-56 GHz), V-band (47-76 GHz) and W-band (71-110 GHz). This paper presented a dedicated data acquisition and control system (DAQC) to meet the measurement requirements with high time and amplitude resolution. The DAQC consisted of two control modules, which included arbitrary waveform generation block (AWG) and trigger processing block (TP), and two data acquisition modules (DAQ) that was implemented base on the PXIe platform from National Instruments (NI). All the performance parameters can satisfy the requirements of reflectometry. The actual performance will be further examined in the experiments of EAST in 2014.

## I. Introduction

Microwave reflectometry is a non-intrusive plasma diagnostic tool which is widely applied in many fusion devices[1-2]. By transmitting microwave beam and collecting echo reflected by the plasma cut-off, reflectometry detects the group delay of electromagnetic waves which can be used to back-calculate the density profile and density fluctuation of plasma. Selection of operating microwave frequency depends on the plasma size, the plasma density and the magnetic field strength. With the increase of the parameters, reflectometry need to work in a broader microwave frequency band[3].

In 2014, the microwave reflectometry on Experimental Advanced Superconducting Tokamak (EAST) [4] had been upgraded. The broad-band microwave reflectometry consisted of 7 channels: 3 channels for the density profile measurement (MRP) that respectively sweep frequency in Q-band (32-56 GHz), V-band (47-76 GHz) and W-band (71-110 GHz); 4 channels for the density fluctuation measurement (MRF) operating at fixed frequency in V-band (47-76 GHz).

TABLE I. CHANNELS OF BROAD-BAND MICROWAVE REFLECTOMETRY

| Channel Name | Band | Mode | Frequency Sweep? | Target |
|---|---|---|---|---|
| **MRP_1** | Q | X | Y | Profile |
| **MPR_2** | V | X | Y | Profile |
| **MRP_3** | W | X | Y | Profile |
| **MRF_1** | V | O | N | Fluctuation |
| **MRF_2** | V | X | N | Fluctuation |
| **MRF_3** | V | O | N | Fluctuation |
| **MRF_4** | V | X | N | Fluctuation |

To meet the requirements of the microwave reflectometry, a dedicated data acquisition and control system (DAQC) was developed. There were three aspects. First, 14-channel I/Q signals from 7 reflectometry channels should be continuously digitized for hundreds of seconds with high accuracy and time resolution. Then obtained mass data had to be stored in real time. Secondly, the voltage-controlled oscillators (VCO) in MRP needed to be controlled by 3-channel voltage signals with high amplitude and low noise to sweep frequency. Meanwhile, the frequency synthesizers in MRF should be set via general-purpose interface bus (GPIB). At last, the trigger signal from the control center of EAST had to be distributed to every digitizer in DAQC for synchronization.

## II. Architecture

As shown in Fig. 1. , the DAQC included a remote controller, a data warehouse, two control modules and two

Manuscript received May 22, 2014. This work is supported by the National Magnetic Confinement Fusion Program of China (No.2014GB106003) and National Natural Science Foundation of China (Nos.11275234, 11305215，11305208).

Fei Wen, Tao Zhang, Shoubiao Zhang, Defeng Kong, Yuming Wang, Xiang Han, Hao Qu and Xiang Gao are with the Institute of Plasma Physics, Chinese Academy of Sciences, Hefei 230031, China (telephone: +86-551-65590405, e-mail: wenfei @ipp.ac.cn). .

data acquisition modules (DAQ) which respectively were used for MRP and MRF. The remote controller directly controlled the DAQ modules via Ethernet. The DAQ modules acquired the I/Q signals from microwave reflectometry and uploaded data to data warehouse. The control module, which connected to DAQ module via universal serial bus (USB), controls the VCO in MRP to sweep frequency; meanwhile the control module got trigger from EAST control center and distributed the processed triggers to digitizers. The frequency synthesizers in MRF were controlled by the DAQ module by GPIB.

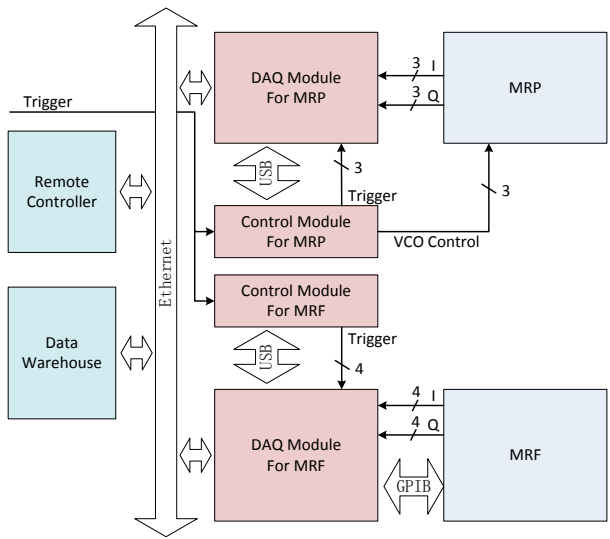

Fig. 1. Block diagram of the data acquisition and control system for broad-band microwave reflectometry.

The DAQ module consisted of high-speed and low-speed digitizers, timing module, controller and redundant arrays of inexpensive disks (RAID), which was based on the PXIe platform from National Instruments (NI). The high-speed digitizer (NI PXIe-5122, 100 MSPS@14 bits) provides two independent inputs to acquire I and Q signal respectively. The number of high-speed digitizers was consistent with the number of channels in reflectometry. The low-speed digitizer (NI PXIe-6368, 2 MSPS) was used to monitor the control signal of VCO in MRP or frequency synthesizer in MPF. The timing module (NI PXIe-6672) generated the sample clock distributed to digitizers by the STAR trigger bus, which ensure the clock reach the digitizers at precisely the same time. The RAID with capacity of 12TB and write speed of 1GB/S could store the data from digitizer in real time.

The control module integrated arbitrary waveform generation block (AWG) and trigger processing block (TP). The AWG generate the VCO control signal according to the waveform data and period parameter given by the user. Before outputting, the generated signals was amplified to the amplitude rang of 0-20V to meet the input dynamic range of VCO. The TP received the trigger from control center and generated a series of triggers to digitizer to complete the given DAQ task.

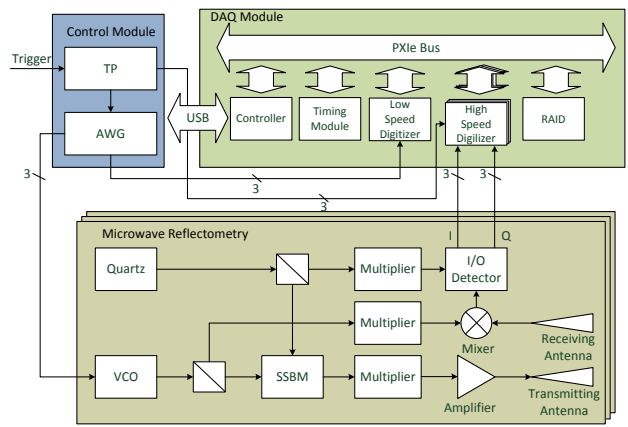

Fig. 2. Block diagram of the control module and DAQ module for MRP.

### III. CONCLUSION

In this paper, we presented a DAQC system for the broad-band microwave reflectometry on EAST. All the performance parameters can satisfy the measurement requirements with high time and amplitude resolution. The actual performance will be further examined in the experiments of EAST in 2014.